\documentclass[10pt,conference]{IEEEtran}

\usepackage{graphicx}
\graphicspath{{figures2/}}
\usepackage{booktabs}
\usepackage{xurl}
\usepackage{balance}

\usepackage{amsmath}

\usepackage{amssymb}

\usepackage{hyperref}
\usepackage{algorithm}
\usepackage{algorithmic}
\usepackage{xcolor}
\usepackage{multirow}

\begin{document}

\title{PIXHELL Attack: Leaking Sensitive Information from Air-Gap Computers via `Singing Pixels'}
%
%


\author{\IEEEauthorblockN{Mordechai Guri}
	\IEEEauthorblockA{Ben-Gurion University of the Negev, Israel}
		Department of Software and Information Systems Engineering\\ Air-Gap Research Lab (\url{http://www.covertchannels.com}) \\ Demo video: \url{https://youtu.be/TtybA7C47SU}  \\
		Email: gurim@post.bgu.ac.il}

\maketitle              

\begin{abstract}	
Air-gapped systems are disconnected from the Internet and other networks because they contain or process sensitive data. However, it is known that attackers can use computer speakers to leak data via sound to circumvent the air-gap defense. To cope with this threat, when highly sensitive data is involved, the prohibition of loudspeakers or audio hardware might be enforced. This measure is known as an `audio gap'.

In this paper, we present PIXHELL, a new type of covert channel attack allowing hackers to leak information via noise generated by the pixels on the screen. No audio hardware or loudspeakers is required. Malware in the air-gap and audio-gap computers generates crafted pixel patterns that produce noise in the frequency range of 0 - 22 kHz. The malicious code exploits the sound generated by coils and capacitors to control the frequencies emanating from the screen. Acoustic signals can encode and transmit sensitive information. We present the adversarial attack model, cover related work, and provide technical background. We discuss bitmap generation and correlated acoustic signals and provide implementation details on the modulation and demodulation process. We evaluated the covert channel on various screens and tested it with different types of information. We also discuss \textit{evasion and stealth} using low-brightness patterns that appear like black, turned-off screens. Finally, we propose a set of countermeasures. Our test shows that with a PIXHELL attack, textual and binary data can be exfiltrated from air-gapped, audio-gapped computers at a distance of 2m via sound modulated from LCD screens.


\end{abstract}

\section{Introduction}
Information security is a major concern for organizations and industries. A wide range of threats, including ransomware, phishing attacks, data breaches, and nation-state offensive activities, characterize the modern cybersecurity environment. To address these emerging challenges, security technologies, threat intelligence, and regulatory frameworks continue to evolve. As part of defenders' efforts to protect sensitive information, they implement robust security measures. They raise awareness about cybersecurity risks, and use a variety of security solutions such as firewalls, data leakage prevention, anomaly detection systems and more.

\subsection{Air-gap Networks}
One of the strategies to protect sensitive and confidential information is the `air-gap' security measure. In this strategy, a network is physically isolated from external networks, meaning there is no direct wired or wireless connection to the Internet \cite{Guri:2018:BAM:3200906.3177230}.
Air-gapped networks are immune to many types of online cyber threats, such as remote exploitation, malware infection, and phishing attacks, as there is no direct connection to external networks. In addition, since there is no direct connection to the outside world, the risk of unauthorized access and data leakage is significantly reduced.

Air-gap networks may be employed when susceptible data is involved. Certain industries, such as healthcare, finance, and defense, may be subject to regulations that mandate air-gapped networks to protect sensitive data and comply with industry standards. For example, stock exchange computer networks may be disconnected from the Internet because they are considered confidential \cite{WaybackM84:online}. Air-gapping may be employed in critical infrastructure sectors such as energy, transportation, and manufacturing to safeguard control systems from cyber threats that could lead to physical harm or disruption.
There may be significant restrictions on maintaining air-gapped networks, such as limiting the use of removable media in the network (USB drives, external hard drives) and preventing LAN or WAN connectivity. Additionally, air-gapped systems may require strict access controls, including biometric authentication and surveillance, to prevent unauthorized access \cite{Beatingt16:online}\cite{clark2009hardware}\cite{guri2021usbculprit}.

\subsection{Air-gap Breaches}
Despite the high level of security and isolation, air-gapped networks are not completely immune to breaches. One of the examples is the Stuxnet worm, which was believed to specifically target supervisory control and data acquisition (SCADA) systems used in Iran's nuclear facilities \cite{karnouskos2011stuxnet}. It spread via infected USB drives and exploited vulnerabilities in the Windows operating system. Agent.btz is another worm that gained notice for its involvement in a significant cybersecurity cyberattack on the United States military's classified and unclassified networks \cite{faou2020agent}. This incident was one of the most serious breaches of U.S. military networks at the time. Agent.btz was primarily spread in disconnected systems via removable media such as USB drives. When an infected USB drive is connected to a computer, the worm copies itself onto that system. In 2018, the U.S. Department of Homeland Security reported that the Russian hacking group compromised the air-gapped systems of America's electric utilities by exploiting third-party vendors in a so-called supply chain attack \cite{America10:online}.
In August 2023, it was reported that a nation-state actor with links to China was suspected of launching a series of attacks last year against industrial organizations in Eastern Europe to siphon data stored on air-gapped systems. The attacks entailed using more than 15 distinct implants and their variants. These implants were broken down into three broad categories based on their ability to establish persistent remote access. They could also gather sensitive information and transmit the collected data to actor-controlled infrastructure \cite{ChinasAP62:online}.

\subsection{Leaking Information from Air-gapped Facilities} 
After breaching the air-gap networks, the attack might want to continue its activity and move on to the subsequent phases of APT attacks, such as lateral movement and exfiltration.
Security firm ESET pointed out more than 15 attack frameworks designed to breach air-gapped networks that were publicly documented, including USBCulprit, USBStealer, Ramsay, PlugX, and others \cite{dorais2021jumping}. As noted in the report, all these malware used USB media to transfer data in and out of the air-gap environment. However, in secured environments, external media may be strictly forbidden or regulated \cite{USBDevic57:online}. In these cases, attackers might exploit `air-gap covert channels' for exfiltration. Using these methods, malware could modulate binary information on top of physical mediums such as electromagnetic emission, optical emanation, and acoustic waves. Researchers have demonstrated that components such as computer RAM, fan noise, keyboard LEDs, and power supply emissions can modulate data \cite{Guri:2018:BAM:3200906.3177230}\cite{carrara2016air}.

One of the main air-gap covert channels explored in the past is acoustic covert channels that use computer speakers. In this method, data is modulated and transmitted from a computer over sound waves generated from the computer's loudspeakers or built-in speakers in the sonic or ultrasonic bands \cite{wong2018crossing}.

\subsection{Air-Gap, Audio-Gap Environments} 
Although acoustic covert channels from loudspeakers were extensively explored in past works \cite{wong2018crossing} it might not always be practical. In secured networks, audio-capable hardware and speakers may not be allowed to create a so-called `audio-gapped' environment \cite{AirGapCo10:online}. Since audio-gap computers lack loudspeakers, the acoustic covert channels described above are impossible. 

\subsection{PIXHELL Attack}
This paper presents the PIXHELL attack, an acoustic covert channel for leaking information from audio-gapped systems. LCD screens contain inductors (coils) and capacitors as part of their internal components and power supply. For example, electrical current passing through coils can cause them to vibrate at an audible frequency, producing high-pitched noise. This phenomenon is known as `coil noise' or `coil whine' \cite{Electrom69:online}. Also, when alternating current (AC) passes through the screen capacitors, they vibrate at specific frequencies. The acoustic emanates are generated by the internal electric part of the LCD screen. Its characteristics are affected by the actual bitmap, pattern, and intensity of pixels projected on the screen. By carefully controlling the pixel patterns shown on our screen, our technique generates certain acoustic waves at specific frequencies from LCD screens.

\subsection{Our Contribution}
Our contribution is as follows.
\begin{itemize}
	\item \textbf{Air-gap, audio-gap attack.}
	We introduce an acoustic convert channel that does not require audio hardware, loudspeaker, or internal speaker on the compromised computer. Instead, we use the LCD screen to generate acoustic signals using designated bitmap patterns.
	
	\item \textbf{Transmission and reception.}
	We designed and implemented a transmitter and receiver and modulation and demodulation algorithms.
	
	\item \textbf{Evasion and stealth.}
	To avoid detection, we used a low-brightness pixel pattern that is difficult to detect. During this attack, the screen may appear dark.
	
	\item \textbf{Multiple transmitters.}
	Our research evaluates the capability of maintaining the covert channel with multiple screen transmitters to increase the bandwidth.
	
	\item \textbf{Splitted patterns.}
	We show that the bitrate can be increased by using split screen techniques. We visualize different pixel patterns concurrently on the screen, enabling modulations such as OFDM.
	
	\item \textbf{Countermeasures.} We discuss countermeasures to this type of acoustic covert channel.
	
\end{itemize}

The paper is organized as follows. We discuss the related work in Section \ref{sec:related}, present the adversarial attack model in Section \ref{sec:adv}, and examine modulation techniques and different types of receivers in Section \ref{sec:trans}. We evaluate the proposed covert channel in Section \ref{sec:eval}, and present countermeasures in Section \ref{sec:cnt}. We conclude in Section \ref{sec:conclude}.

\section{Related Work}
\label{sec:related}
The term air-gap covert channels refers to methods of communication that allow the transfer of information between two systems or networks that are physically separated \cite{carrara2016air}. Researchers and attackers have demonstrated various techniques for bypassing this isolation, creating covert channels for data transfer over air gaps. Air-gap covert channels can be categorized into several groups.

Electromagnetic emanations from electronic devices like computers can be exploited as a covert channel. By modulating electromagnetic radiation, information can be transmitted and received with specialized equipment. Electromagnetic extraction methods have been investigated in the past. AirHopper attack uses the screen cables as antennas to emit radio signals at the FM bands \cite{guri2014airhopper}. GSMem exploited the memory and CPU bus to leak data out of air-gapped computers \cite{guri2015gsmem}. Other works use data exfiltration from memory in various ways and modulation schemes for data exfiltration in covert and side channels, including LoRa \cite{shen2021lora}, BitJabber \cite{zhan2020bitjabber}, and RAMBO \cite{guri2023rambo}.
The operation of optical covert channels involves using light emissions or variations in screen brightness to transmit data. For instance, a malware-infected system could manipulate the display's brightness, and a nearby camera on another system could capture and interpret the variations \cite{guri2019brightness}. Other techniques utilize indicator LEDs such as the keyboard, routers, printers, and LCD screens \cite{guri2019ctrl}\cite{cronin2019covert}. 
Electric covert channels like PowerHammer involve modulating electrical characteristics, such as voltage or current, to transmit data. A simple sensor could capture and interpret these changes \cite{guri2019powerhammer}.
Electronic components generate magnetic fields to encode and transmit data. Magnetic sensors on nearby devices could capture and interpret these variations. Covert channels that exploit magnetic fields, including Odini \cite{guri2019odini}, Magneto \cite{GURI2021115}, MagView \cite{zhang2020magview}, and others \cite{matyunin2016covert}, usually have limited bit rates.
Thermal covert channels leverage temperature variations to transmit data. Malicious software can manipulate CPU and GPU temperatures; a temperature sensor on another PC could interpret these variations. Few thermal covert channels have been successfully demonstrated between computers \cite{guri2015bitwhisper} and cores \cite{masti2015thermal}.

\subsection{Acoustic Communication in Air-Gapped Systems}
Acoustic communication channels use sound or ultrasonic frequencies to enable data transfer between isolated systems \cite{wong2018crossing}. This method typically involves malware on a secured system manipulating speaker outputs to generate sounds encoded with data. A microphone on a nearby device can capture these sounds, decoding them to retrieve the original data. Previous research has largely focused on the capabilities of loudspeakers in these covert operations. Studies by Carrara \cite{carrara2015acoustic} and Hanspach \cite{hanspach2014covert} delved into the communication capabilities and characteristics of air-gapped systems, exploring various scenarios and practical applications of these covert channels. Hanspach et al. have successfully shown that it's possible to send data through the air using ultrasonic frequencies undetectable by human ears, highlighting the technique's potential for creating a mesh network and its security concerns \cite{hanspach2014covert}. Beyond computers, acoustic signals have been used to transmit data between mobile devices \cite{pandya2022shoutimei}\cite{deshotels2014inaudible}. Other attacks, such as the MOSQUITO \cite{guri2018mosquito} and SpeakEar \cite{Guri2017e}, have even transformed computer speakers into microphones. Recent advancements include Sherry et al.'s demonstration of a software-defined radio (SDR) approach for establishing ultrasonic frequency channels with low bandwidth \cite{sherry2023near}, and Zhang et al.'s discussion on ultrasound-based communication among smart devices \cite{zhang2022ultrannel}. Techniques like AirViber \cite{guri2021exfiltrating} and Gairoscope \cite{guri2021gairoscope} leverage mechanical vibrations from computer parts for data transmission, with smartphones acting as receivers. Additionally, Matyunin has introduced a method using vibrations from low-frequency acoustic signals for covert communication \cite{matyunin2019vibrational}.

\subsection{Overcoming Audio-Gap Restrictions}
Researchers have developed methods of exfiltrating sound from computer systems that lack speakers as a result of audio-gapping security measures. Guri et al. introduced Fansmitter, a technique that varies the noise of computer fans (CPU/GPU) through malware, with these sound variations being detectable by nearby smartphones \cite{guri2020fansmitter}\cite{guri2022gpu}. Diskfiltration leverages hard disk drive noise \cite{guri2017acoustic}, while CD-LEAK uses sounds from CD/DVD drives for data modulation \cite{guri2020cd}. The power-supply attack, introduced by Guri et al. in 2020, demonstrates how power supply units can be manipulated to emit sound or ultrasonic waves without traditional speakers \cite{guri2021power}. More recently, inkjet printers have been adapted by Briseno et al. to send mechanical acoustic signals, allowing for the low-rate transfer of sensitive data over short distances \cite{de2022inkfiltration}.

\section{Attack Model}
\label{sec:attackmddel}
The attack model on air-gapped networks is composed of three main phases: (1) network infiltration, (2) data gathering, and (3) data exfiltration.

{\textbf{Network infiltraton.} In computer security, an air gap is a measure to ensure that a secure network is completely disconnected from unsecured networks, including the Internet. However, despite the isolation, determined attackers may still find ways to breach the air gap and compromise the security of the isolated system, installing high-profile malware or APT. Attackers may gain physical access to the isolated system through direct infiltration or by manipulating insiders. Once physical access is achieved, malware or malicious hardware can be introduced to compromise the system \cite{dorais2021jumping}. USB drives or other removable media can be used as malware delivery vectors. An attacker may infect a device outside the air-gapped environment, insert an infected USB drive into the isolated system, and execute or transfer malware. Phishing, malicious insiders, or other social engineering techniques may be employed to trick individuals with access to the air-gapped system into taking actions that compromise security, such as clicking on malicious links or downloading infected files. Attackers may also use software supply chain attacks by targeting software application dependencies or third-party libraries. By compromising these dependencies, they can introduce vulnerabilities or malicious code that may go unnoticed during development and testing \cite{martinez2021software}.

{\textbf{Data gathering.} At the second stage, the malware may gather information of interest, including files, keylogging, biometric information, encryption keys, images, etc. The information might be collected locally or by several instances of the APT and may be kept in a persistent way in a file system on the hard disk drive.
		
{\textbf{Data exfiltration.}
As part of the third phase of the attack, the APT may choose to exfiltrate the information. At this stage, the data is encoded and exfiltrated acoustically, modulated over the acoustic signals emanating from the LCD screens of local computers. Nearby microphones, compromised laptops, or malicious-infected smartphones can collect acoustic signals. The receiving device records the acoustic signals, demonstrates and decodes them, and then sends them to the attack over the Internet.

Figure \ref{fig:illus} illustrates the attack scenario. A malware infection on the compromised computer (A) encodes information and uses crafted pixel patterns to exfiltrate it over the emanated acoustic signals. A nearby laptop computer receives the signals, decodes them, and sends them to the attacker.

\begin{figure}
	\centering
	\includegraphics[width=\linewidth]{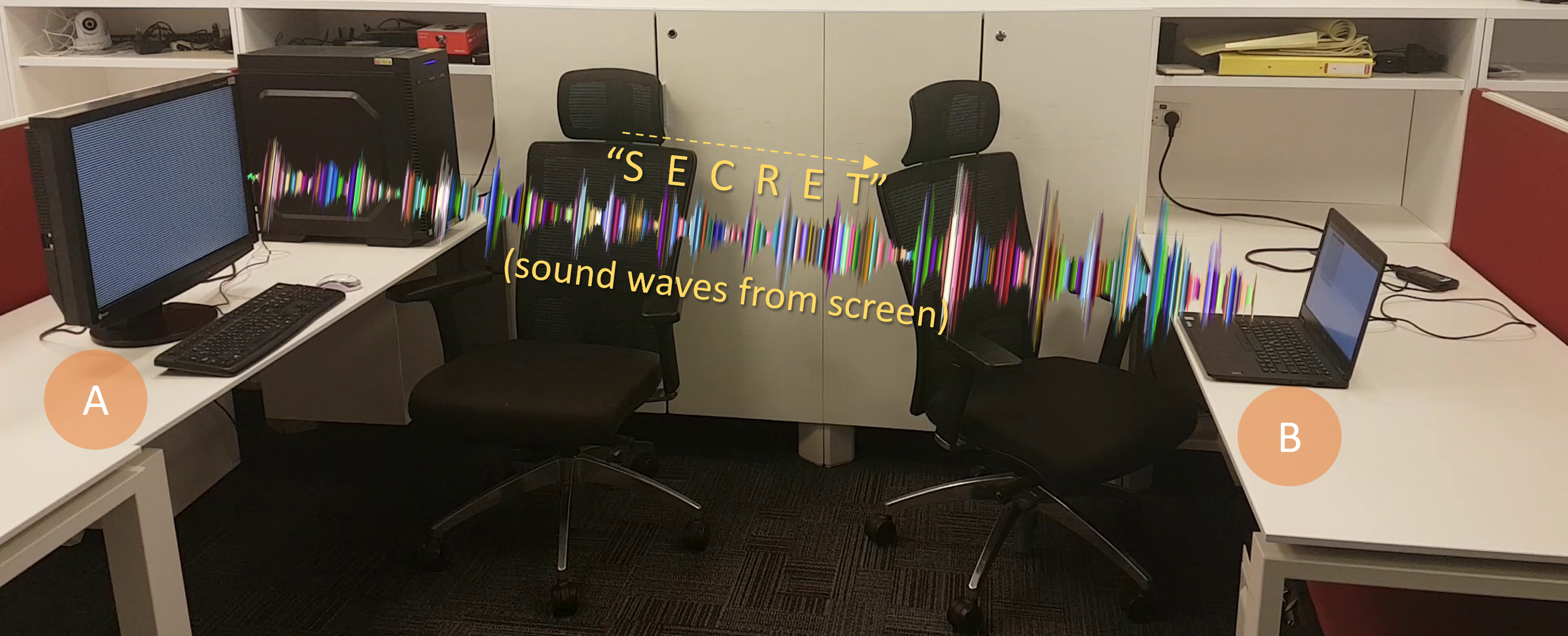}
	\caption{Attack scenario. A malware infection on the compromised computer (A) encodes information and uses crafted pixel patterns to exfiltrate it over the emanated acoustic signals. A nearby laptop computer receives the signals, decodes them, and sends them to the attacker.}
	\label{fig:illus}
\end{figure}

\section{Technical Background}
\label{sec:adv}
\subsection{LCD Screen}
An LCD screen is built from several distinct layers, each playing a crucial role in displaying images. At the forefront, a polarizing layer filters and polarizes the light that enters the screen, allowing only light of a specific polarization to proceed. Beneath this, glass substrates house thin-film transistors (TFTs), which are microscopic electronic switches. These switches manage the activation of individual pixels across the screen. Functioning as semiconductor devices, TFTs organize into a grid, with each pixel linked to its own transistor. These transistors regulate the electrical current flowing to the layer of liquid crystals. Directly above the transistors, color filters are embedded, consisting of red, green, and blue hues. These filters assign the color of each pixel. When an electric current is applied, the liquid crystals adjust their orientation, thus altering their optical characteristics. This change allows light to either pass through or be blocked, depending on the voltage applied to each pixel. Behind the layer of liquid crystals, another polarizing layer is placed, ensuring that only correctly oriented light can reach the viewer's eyes. For visibility, LCD screens require a backlight, typically provided by cold cathode fluorescent lamps (CCFL) or LEDs in contemporary models. This backlight emits a consistent stream of light that traverses the screen's layers to produce visible images. A diffuser layer sits between the backlight and the liquid crystals, ensuring the light spreads uniformly across the display.

\subsection{The Acoustic Effects in LCD}
Acoustic noise from LCD screens results from the dynamic interaction between bitmap patterns and power supply and capacitor loads, resulting in variations in acoustic emissions. These changes in power consumption can cause mechanical vibrations or piezoelectric effects in capacitors, producing audible noise. Coils within the power supply, known as inductors, can also contribute to noise through a phenomenon known as coil whine. In `coil whine', electrical current flows through the coil, causing it to vibrate at audible frequencies, particularly during periods of varying loads induced by varying screen images. The correlation between specific bitmap patterns and the acoustic frequency generated from a screen hinges on how these patterns influence the electrical behavior of the screen's components, notably the power supply capacitors and inductors (coils). This leads to variations in physical phenomena like coil whine or the piezoelectric effect, which generate sound.

In the context of the PIXHELL attack, certain bitmap patterns require more power to display than others. Particularly, a white pixel demands more power from the supply than a dark pixel. Intermediate patterns vary in power consumption based on the number of pixels lit and their distribution across the screen. Consequentially, electronic components within the LCD monitor, such as transformers or conductors, can produce high-pitched noise when they vibrate at specific frequencies, reflecting the current bitmap pattern shown on the screen.
\subsection{Other LCD Noise Sources}
Notably, LCD screens have additional sources of noise. The backlight inverter, which powers the LCD screen backlight, may produce a faint whining noise. This can occur due to the inverter's electrical components. The LCD monitor may pick up interference from other electronic devices or cables. This interference can manifest as a whining sound through the monitor's built-in speakers or as electromagnetic interference, causing speakers near emit noise. The inverter can produce audible high-frequency noise due to rapid voltage switching. Integrated circuits used for voltage regulation can generate high-frequency noise as they maintain stable voltage levels for various components within the monitor.

\section{Transmission}
This section discusses the signal generation, modulation scheme, and communication protocol.
\label{sec:trans}
\subsection{Signal Generation}
The acoustic signal is generated when creating black-and-white rows on the screen. The width of the rows corresponds to the frequency of the audio transmission.
The pixel clock frequency is measured in Hz and represents the number of pixels transmitted per second. It is calculated by multiplying the total number of pixels in each row by the total number of rows (vertical resolution) and then multiplying by the refresh rate, as shown in Equation \ref{eq:1}.

\begin{equation}
	\label{eq:1}
	\centering
	P_{clk} = H_{res} \times V_{res} \times R_{rates}
\end{equation}

The modulation of the bitmap on the monitor is defined by several parameters, including the pixel clock frequency \( P_{\mathrm{clk}} \), the horizontal resolution \( H_{\mathrm{res}} \), the vertical resolution \( V_{\mathrm{res}} \), and the total number of pixels per cycle. The total pixel count per cycle can be expressed as:

\[
\text{cycleSize} = \frac{P_{\mathrm{clk}}}{f}
\]

where \( P_{\mathrm{clk}} \) is the pixel clock frequency and \( f \) is the modulation frequency. The cycle is divided into two equal halves:

\[
\text{halfCycle} = \frac{\text{cycleSize}}{2}
\]

Given these parameters, the modulation process computes the pixel positions based on the horizontal and vertical coordinates, denoted by \( x \in [0, H_{\mathrm{res}}] \) and \( y \in [0, V_{\mathrm{res}}] \).

For each pixel located at \( (x, y) \), a sample number is calculated as:

\[
\text{sampleNumber} = x + y \cdot H_{\mathrm{total}}
\]

where \( H_{\mathrm{total}} \) represents the total horizontal pixel count, including blanking intervals. The modulation is determined by the position of the sample number within the modulation cycle. Specifically, the remainder of the sample number, when divided by the cycle size, is used to decide the pixel state:

\[
\text{remainder} = \text{sampleNumber} \mod \text{cycleSize}
\]

The pixel value is assigned based on the remainder. If \( \text{remainder} < \text{halfCycle} \), the pixel value is set to white \( (0xFF) \). Otherwise, the pixel value is set to black \( (0x00) \). This modulation scheme produces a visual pattern of alternating black and white regions that corresponds to the modulation frequency.

This process is applied to each pixel in the display. By iterating over the full screen resolution, the modulation creates a binary signal where the display alternates between black and white strips. These strips are aligned with the square wave defined by the modulation frequency, resulting in a visually modulated output. The overall process generates a modulated visual representation on the screen, using precise pixel placement based on the calculated cycle parameters.

Figure \ref{fig:tones} shows four bitmap patterns generated on the screen with the signal generation algorithm, modulating an acoustic signal at 5000 Hz, 10000 Hz, 15000 Hz, and 20000 Hz.

\begin{figure*}
	\centering
	\includegraphics[width=\textwidth]{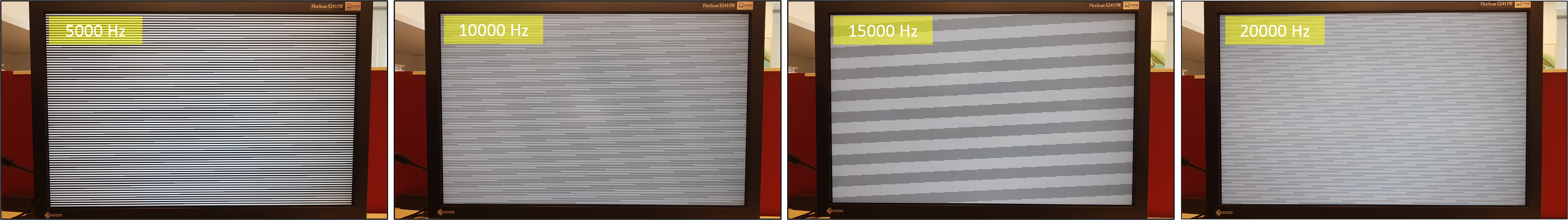}
	\caption{Four bitmap patterns generated on the screen with the signal generation algorithm, modulating an acoustic signal at 5000 Hz, 10000 Hz, 15000 Hz, and 20000 Hz.
	}
	\label{fig:tones}
\end{figure*}

\subsection{Modulation}
The signal generation technique allow the implementation of three basic types of modulations: On Off Keyking (OOK), Frequency Shift Keying (FSK), and Amplitude Shift Keying (ASK).

\subsubsection{On-Off Keying (OOK)}
On-Off Keying (OOK) is a simple modulation where the presence or absence of a carrier wave represents binary data. In OOK, binary '1' is represented by the presence of a carrier wave at a specific frequency, while binary '0' is represented by the absence of the carrier wave. In our case, a bitmap produced at a given frequency is a carrier, which is then modulated according to the transmitted bits. It is expressed as
\[
s_{\text{OOK}}(t) = A \cdot b(t) \cdot \cos(2 \pi f_c t + \phi)
\]

where $A$ is the amplitude of the carrier, $b(t)$ is the binary data signal 1 for the presence of the carrier and 0 for its absence, $f_c$ is the carrier frequency, $\phi$ is the initial phase of the carrier, and $t$ is time.

\subsubsection{Frequency Shift Keying (FSK)}
Frequency Shift Keying (FSK) is a modulation scheme in which the frequency of a carrier wave is varied in accordance with the binary data being sent. This method encodes binary data into the frequency variations of a continuous carrier wave, with different frequencies representing different binary values. The simplest form of FSK is Binary FSK, where only two frequencies are used where one frequency ($f_{1}$) represents a binary '1', and another frequency ($f_{2}$) represents a binary '0'. MFSK uses more than two frequencies, allowing it to encode more bits per symbol. It is expressed as
\[
s_{\text{FSK}}(t) = A \cdot \cos(2 \pi f(t) t + \phi)
\]

where $A$ is the constant amplitude of the carrier, $f(t)$ is the instantaneous frequency of the carrier, which switches between $f_1$ for binary '1' and $f_2$ for binary '0', $\phi$ is the initial phase of the carrier, and $t$ is time.

\subsubsection{Amplitude Shift Keying (ASK)}
Amplitude Shift Keying (ASK) modulation is a form of digital modulation where the amplitude of a carrier wave is varied in accordance with the binary data being transmitted. In ASK, the carrier signal's amplitude switches between two levels, representing binary data. It is expressed as
\[
s_{\text{ASK}}(t) = A(t) \cdot \cos(2 \pi f_c t + \phi)
\]

where $A(t)$ is the time-varying amplitude of the carrier, which changes with the binary data signal (higher amplitude for binary '1' and lower or zero amplitude for binary '0', $f_c$ is the carrier frequency, $\phi$ is the initial phase of the carrier, and $t$ is time.

We implemented each of the three modulation schemes for the PIXHELL transmitter application. 

\subsection{Transmission Protocol}
We used a basic packet structure as follows:
\textbf{Preamble Sequence of Bit.} This part of the header is used for synchronization, allowing the receiver to synchronize its buffer with the sender's clock based on the pattern of bits. The length of this sequence is eight alternating bits (\texttt{0xAA}) and varies depending on the specific protocol. It consists of a repeating pattern that allows the receiver to detect and use timing and amplitude parameters.
\textbf{Payload of 32 Bits.} The payload carries the actual data being transmitted. In this case, it is specified as 32 bits long. This part of the packet contains meaningful information the sender wishes to communicate with the receiver. The payload size can impact the efficiency and speed of data transmission, with 32 bits being relatively small, making it suitable for simple or compact data structures, such as keylogging, texts, encryption keys, etc.
\textbf{CRC of 8 Bits.} CRC (Cyclic Redundancy Check) is used to check for errors in the transmitted packet. The CRC is calculated based on the payload data before transmission. The receiver then recalculates the CRC from the received data and compares it to the transmitted CRC to check for integrity. An 8-bit CRC can detect common types of errors, such as single/double-bit errors, making it a compact and effective choice for error detection in our packets.

Figure \ref{fig:packet} shows the spectrogram of a packet received by a nearby smartphone.

\begin{figure}
	\centering
	\includegraphics[width=0.8\linewidth]{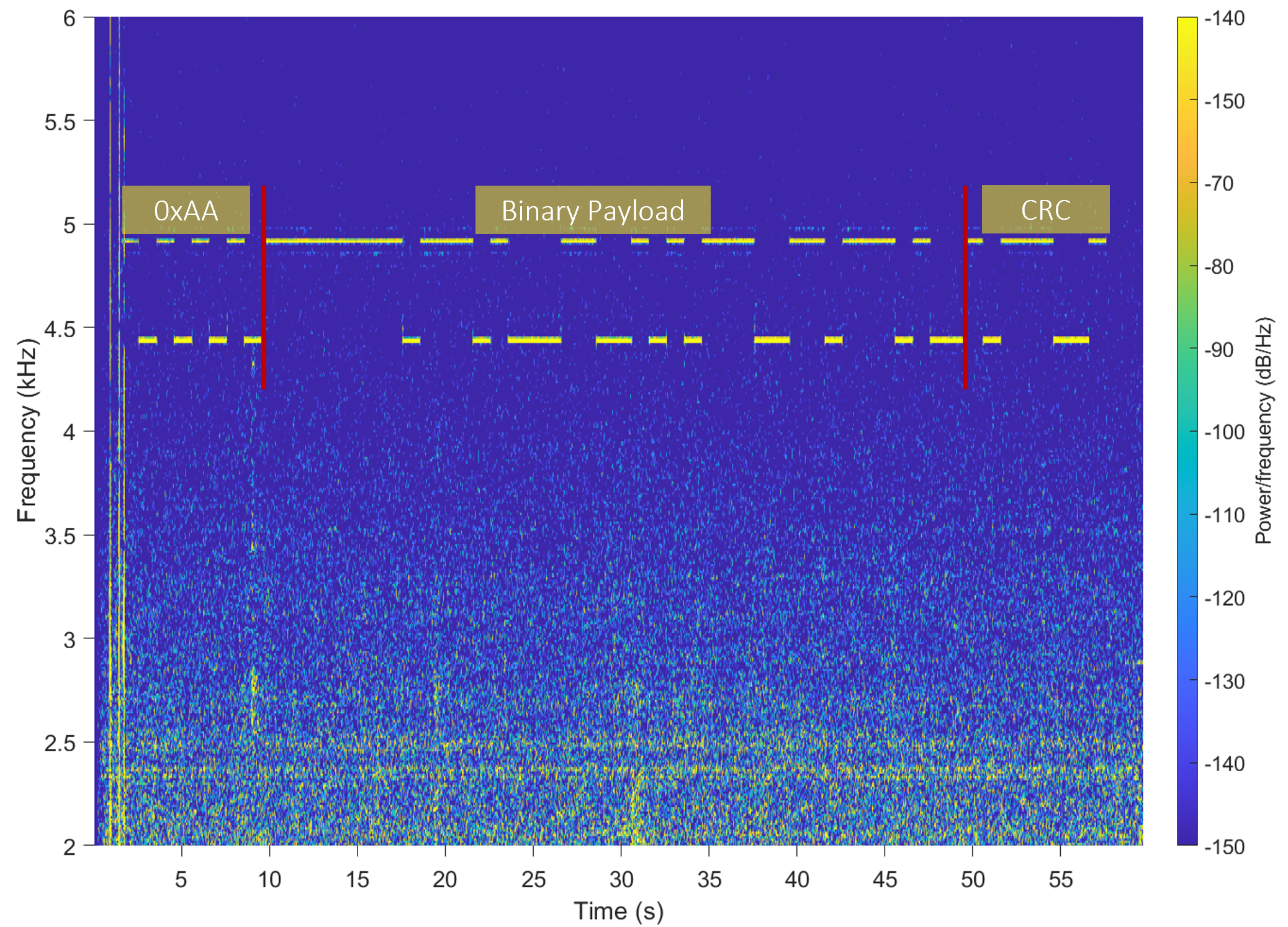}
	\caption{Spectrogram of a packet received by a nearby smartphone.}
	\label{fig:packet}
\end{figure}

\section{Reception}
\label{sec:rec}
We implemented the receiver for Microsoft Windows and Android OS with a similar main algorithm structure. 
The demodulator for the described packet involves several steps presented in Algorithm \ref{alg:demodulatePacket}, focusing on synchronizing with the preamble sequence, extracting the payload, and verifying the integrity of the data using the CRC.

\begin{algorithm}
	\caption{Demodulate Packet}
	\label{alg:demodulatePacket}
	\begin{algorithmic}[1]
		\STATE \textcolor{blue}{\textbf{Input:}} inputStream
		\STATE \textcolor{blue}{\textbf{Output:}} Decoded data
		\STATE \textbf{const} \textbf{PREAMBLE\_PATTERN} = ``10101010''
		\STATE \textbf{const} \textbf{PAYLOAD\_LENGTH} = 32
		\STATE \textbf{const} \textbf{CRC\_LENGTH} = 8
		\WHILE{\text{not end of inputStream}}
		\IF{\text{detectPreamble(inputStream, PREAMBLE\_PATTERN)}}
		\STATE \text{synchronizeWithPreamble(PREAMBLE\_PATTERN)}
		\STATE \textit{payload} = \textit{readBits(inputStream, PAYLOAD\_LENGTH)}
		\STATE \textit{receivedCRC} = \textit{readBits(inputStream, CRC\_LENGTH)}
		\STATE \textit{calculatedCRC} = \textit{calculateCRC(payload)}
		\IF{\text{receivedCRC == calculatedCRC}}
		\STATE \text{processPayload(payload)}
		\ELSE
		\STATE \text{handleCorruptedPacket()}
		\ENDIF
		\ENDIF
		\ENDWHILE
	\end{algorithmic}
\end{algorithm}

The demodulator receives the time-domain signal, which may contain frequencies modulated from the transmitted data. We apply a window function to the signal segment to reduce spectral leakage before FFT. The signal is then processed through an FFT algorithm, transforming it into a spectrum representing the signal's frequency components and their amplitudes in a time window, representing it as a stream. We implemented the OOK, M-FSK, and ASK demodulator versions. For simplicity, the FFT process is not shown in the pseudocode, and we focus instead on the demodulator logic.

The demodulator continuously monitors the incoming bit stream to detect the preamble sequence. This involves a correlation process, where the demodulator compares the incoming bits with the expected preamble pattern. The function \texttt{detectPreamble()} implements correlation or pattern matching to determine the preamble (line 7). When the correlation exceeds a certain threshold, indicating a high probability of preamble detection, the demodulator concludes that the packet start has been found. The function \texttt{synchronizeWithPreamble()} adjusts the bit synchronization based on the detected preamble for accurate bit sampling and extracts the channel parameters, depending on the modulation (ASK, FSK, etc.). After synchronization, the demodulator samples the next 32 bits as the payload of the packet (\texttt{readBits()}) which are read and temporarily stored for further processing (line 10). 
The demodulator calculates the CRC of the extracted payload using the CRC algorithm as the sender. This involves processing the 32 bits of the payload through a predetermined polynomial equation to produce an 8-bit CRC value. If the two values match, the packet is considered error-free, and the payload is passed on to the next stage of processing (\texttt{processPayload()})). If the CRC values do not match, it indicates that the packet has been corrupted during transmission, and the packet is discarded (\texttt{handleCorruptedPacket()}).

Note that since the PIXHELL covert channel is unidirectional, using error correction might be a better choice for the covert channel. Unlike error detection, which merely identifies the presence of errors, error correction mechanisms can identify and fix errors within corrupted data. This ensures the integrity and accuracy of the received or retrieved information. Implementing error correction is possible with various approaches such as Hamming Distance, Forward Error Correction (FEC), and others.
\section{Evaluation}
\label{sec:eval}
We evaluated the covert channel for various parameters described in this section. For the primary tests, we used the four LCD screens listed in Table \ref{tab:LCD}. LCD1 is a 22-inch ViewSonic VA2232WM-LED with a resolution of 1680 x 1050, LCD2 is a 22-inch Samsung SyncMaster 226BW with a resolution of 1680 x 1050, LCD3 is a 24.1-inch Eizo FlexScan with a resolution of 1920 x 1200, and LCD4 is a 40-inch Samsung TV model UA40B6000VRXSQ with a resolution of 1920 x 1080. Our Android app was installed on the Samsung Galaxy S11 phone, and the demodulator was run on the Dell Latitude with Microsoft Windows. Figure \ref{fig:chirp} shows the sweep signal generated from the four screens. As can be seen, the signal wraps around the low frequencies (3-5 kHz) up to 15 kHz and reaches the near ultrasonic band (20 kHz and above) in LCD1 and LCD4.

\begin{table}[]
	\centering
	\caption{Tested LCD screens}
	\label{tab:LCD}
		\begin{tabular}{@{}llll@{}}
			\toprule
			& Screen & Model & Size \\ \midrule
			LCD1 & ViewSonic & VA2232WM-LED & 22-inch 1680 x 1050 \\
			LCD2 & Samsung & syncMaster 226BW & 22-inche 1680 x 1050  \\
			LCD3 & Eizo & FlexScan & 24.1-inch 1920 x 1200 \\
			LCD4 & Samsung TV & UA40B6000VRXSQ & 40-inch 1920 X 1080 \\ \bottomrule
		\end{tabular}%
\end{table}

\begin{figure}
	\centering
	\includegraphics[width=\linewidth]{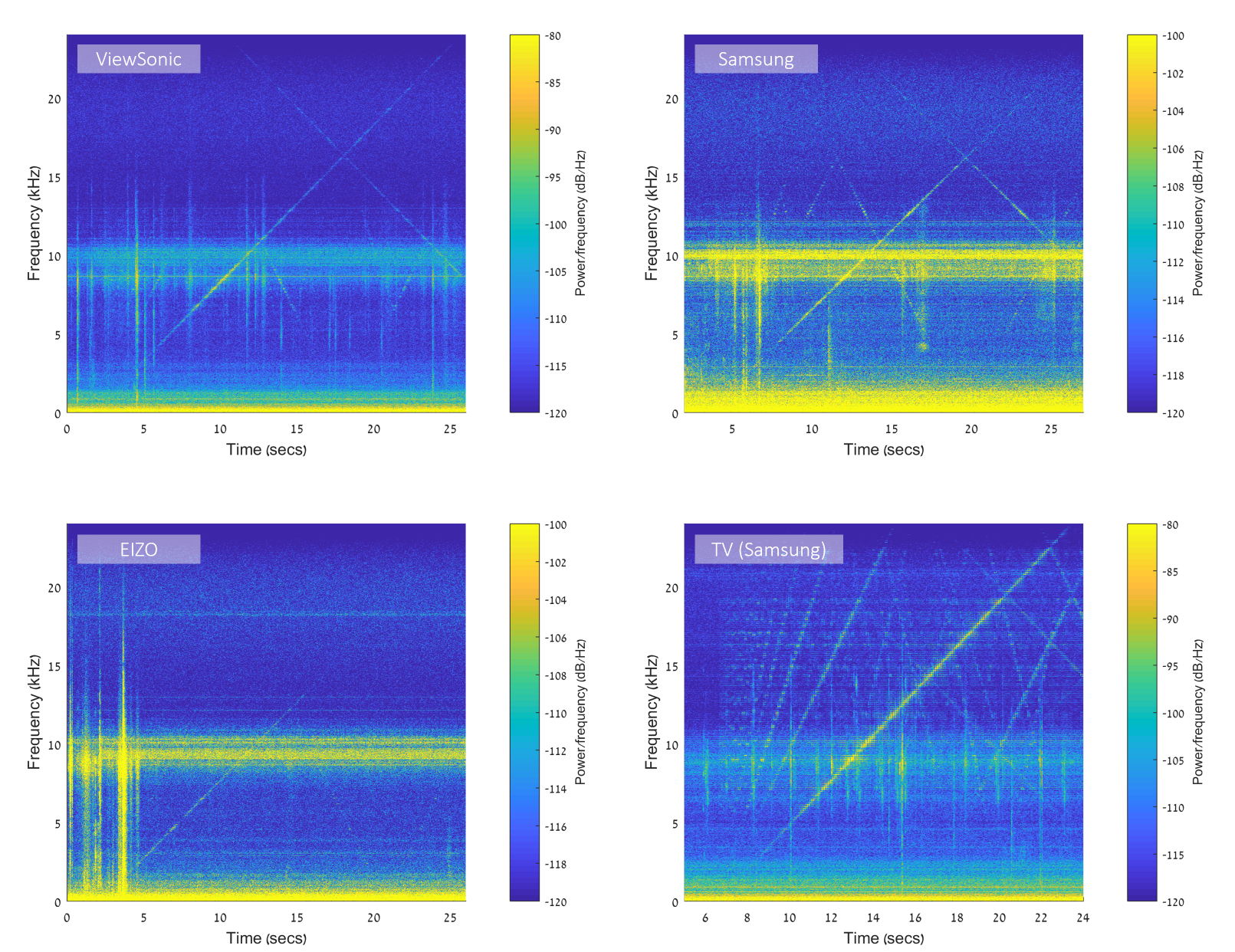}
	\caption{Chirp acoustic signal from four LCD screens}
	\label{fig:chirp}
\end{figure}

\subsection{SNR}
Table \ref{tab:SNR1} summarizes the Signal-to-Noise Ratio (SNR) values for four different display brands (ViewSonic, Samsung, EIZO, and TV) at various distances (from 0 to 2.5 meters) and for three different bit rates (5, 10, and 20 bps). The SNR values across all brands and conditions range broadly from 5 dB (Samsung at 2.5 meters, 20 bps) to 43 dB (TV at 0.5 meters, 5 bps), indicating significant variability based on brand, distance, and bit rate. The TV brand consistently shows higher SNR values at shorter distances (0 and 0.5 meters) across all bit rates, with the peak being 43 dB at 0.5 meters for 5 bps, indicating superior performance in minimizing noise at closer ranges. ViewSonic shows a relatively stable performance across distances and bit rates, with SNR values mostly above 20 dB, peaking at 34 dB at 1 meter for 20 bps.
Samsung demonstrates more variability, particularly struggling at longer distances (2 and 2.5 meters) with SNR dropping to as low as 5 dB at 2.5 meters for 20 bps.
EIZO exhibits the most fluctuation, with low SNR values at 0.5 and 1.5 meters across all bit rates, but recovers at 1 and 2 meters, suggesting inconsistency in performance based on distance.
TV stands out for maintaining high SNR values at closer distances, though it shows a decline as distance increases, maintaining better performance than the other brands at longer distances.
Note that There's a general trend where SNR decreases as distance increases, which is expected due to signal attenuation. However, the rate of decline varies by brand and bit rate, with some brands maintaining better performance over distance.

\begin{table*}[]
	\centering
	\caption{Measured SNR}
	\label{tab:SNR1}
	\resizebox{\textwidth}{!}{%
		\begin{tabular}{@{}lllllllllllll@{}}
			\toprule
			Distance (m) & \multicolumn{3}{c}{ViewSonic} & \multicolumn{3}{c}{Samsung} & \multicolumn{3}{c}{EIZO} & \multicolumn{3}{c}{TV} \\
			& SNR 5 bps & SNR 10 bps & SNR 20 bps & SNR 5 bps & SNR 10 bps & SNR 20 bps & SNR 5 bps & SNR 10 bps & SNR 20 bps & SNR 5 bps & SNR 10 bps & SNR 20 bps \\ \midrule
			0 & 24 dB & 28 dB & 27 dB & 23 dB & 23 dB & 24 dB & 22 dB & 22 dB & 20 dB & 32 dB & 38 dB & 25 dB \\
			0.5 & 26 dB & 23 dB & 26 dB & 20 dB & 22 dB & 20 dB & 12 dB & 11 dB & 11 dB & 30 dB & 38 dB & 36 dB \\
			1 & 27 dB & 30 dB & 34 dB & 21 dB & 19 dB & 20 dB & 23 dB & 22 dB & 21 dB & 32 dB & 30 dB & 22 dB \\
			1.5 & 26 dB & 29 dB & 31 dB & 17 dB & 19 dB & 18 dB & 12 dB & 12 dB & 13 dB & 23 dB & 22 dB & 21 dB \\
			2 & 18 dB & 23 dB & 27 dB & 14 dB & 17 dB & 13 dB & 17 dB & 14 dB & 13 dB & 23 dB & 20 dB & 18 dB \\
			2.5 & 27 dB & 20 dB & 23 dB & 15 dB & 13 dB & 5 dB & - & - & - & - & - &  \\
			\bottomrule
		\end{tabular}%
	}
\end{table*}

\subsection{Location}
The power and quality of the emanated acoustic signal depend on the specific screen structure, its internal power supply, coil and capacitor locations, etc. Table \ref{tab:loc} provides a comparison of signal-to-noise ratio (SNR) values for different screens, with a smartphone receiver positioned on various sides of each screen (back, front, left, right, and on the desk). Each screen is identified by its brand, specific model, and resolution, with all but one model featuring a resolution of 1920x1080 pixels; the Dell E198FPb model stands out with a resolution of 1280x1024 pixels. The SNR values are presented in dB (decibels), indicating the signal strength level relative to background noise from different orientations relative to the screen. Dell E2216HV shows a notably high SNR value of 23.79 dB when the receiver is located at the back. This is significantly better than other screens in similar positions. The ViewSonic VX2370S-LED and BenQ GW2255 exhibit higher SNR values when the receiver is on the desk, at 17.71 dB and 19.16 dB, respectively. This suggests these models perform well in a typical desk setup. HP 2310ti has the highest SNR value of 12.91 dB when the receiver is placed on the desk. This indicates improved performance in this specific setup. The SNR values vary significantly based on the receiver's position relative to the screen, with generally higher values at the back, suggesting that facing the screen from the back often results in better signal reception.

\begin{table*}[]
	\centering
	\label{tab:loc}
	\caption{SNR with a smartphone receiver located at various sides of the screen}
	\begin{tabular}{@{}lllllllll@{}}
		\toprule
		\# & Screen                 & Model     & Resolution & Back    & Front   & Left    & Right   & Desk    \\ \midrule
		1  & ViewSonic VA2232WM-LED & VS14517   & 1920x1080  & 8.84 dB  & 10.08 dB  & 8.7 dB   & 6.52 dB  & 9.34 dB  \\
		2  & HP 2310ti              & X         & 1920x1080  & 8.59 dB  & 11.03 dB & 9.98 dB  & 7.44 dB & 12.91 dB \\
		3  & Dell                   & E2216HV   & 1920x1080  & 23.79 dB & 19.25 dB  & 16.07 dB & 12.96 dB & 14.54 dB \\
		4  & ViewSonic VX2370S-LED  & VS14880   & 1920x1080  & 10.80 dB & 12.00 dB  & 12.59 dB & 18.21 dB & 17.71 dB \\
		5  & BenQ GW2255            & GL2250-T  & 1920x1080  & 16.22 dB & 15.28 dB & 12.51 dB & 12.12 dB  & 19.16 dB \\
		6  & Dell                   & E198FPb   & 1280x1024  & 15.94 dB & 8.52 dB  & 9.13 dB  & 8.36 dB  & 12.56 dB \\
		7  & Samsung                & S22A450MW & 1920x1080  & 19.17 dB & 9.58 dB  & 8.29 dB  & 13.78 dB  & 10.23 dB \\
		10 & Dell                   & P2419H    & 1920x1080  & 14.96 dB & 19.14 dB & 9.96 dB  & 9.06 dB & 17.98 dB \\
		12 & Samsung                & TV        & 1920x1080  & 18.22 dB & 9.5 dB   & 8.87 dB   & 7.6 dB  & 13.60 dB \\ \bottomrule
	\end{tabular}
\end{table*}

\subsection{Orthogonal Frequency-Division Multiplexing (OFDM)}
In Orthogonal Frequency-Division Multiplexing (OFDM), the signal is divided into sub-carriers maintained simultaneously. By splitting the screen into $n$ strips, it is possible to implement OFDM modulation. This can be done by dividing a data signal across multiple closely spaced carrier frequencies to provide efficient data transmission. A portion of the screen modulates each carrier frequency. Because they are orthogonal, they can be tightly packed without interfering with each other. This orthogonality allows efficient spectrum utilization. The bitmap projected on the screen encodes $n$ different frequencies. Figure \ref{fig:multi} shows the spectrogram of FSK (left) and OFDM (right) modulated acoustically from an LCD screen. In this case, two frequencies are demonstrated where $f_{a}$ and $f_{b}$ are generated by two encoded bitmaps (Figure \ref{fig:multi2}). However, it's worthwhile to note that the amount of energy (emitted noise) in each sub-carrier is highly correlated with the size of the bitmap patterns. With the image split into $n$ stripes, the signal strength in each sub-carrier would be divided by $n$. 

Note that selecting appropriate sub-carrier frequencies is essential for maintaining orthogonality and system efficiency. The frequency of the $n$-th sub-carrier ($f_n$) in an OFDM system is determined by the formula $f_n = f_c + n \cdot \Delta f$, where $f_c$ is the central frequency of the OFDM channel, $n$ is the sub-carrier index, and $\Delta f$ is the sub-carrier spacing. The sub-carrier spacing, which is crucial for ensuring orthogonality among the sub-carriers, is chosen as the reciprocal of the system's symbol duration ($T$), i.e., $\Delta f = \frac{1}{T}$. This selection ensures that the sub-carriers are orthogonal over each symbol period, preventing inter-carrier interference and allowing for efficient spectrum use. 

\begin{figure}
	\centering
	\includegraphics[width=\linewidth]{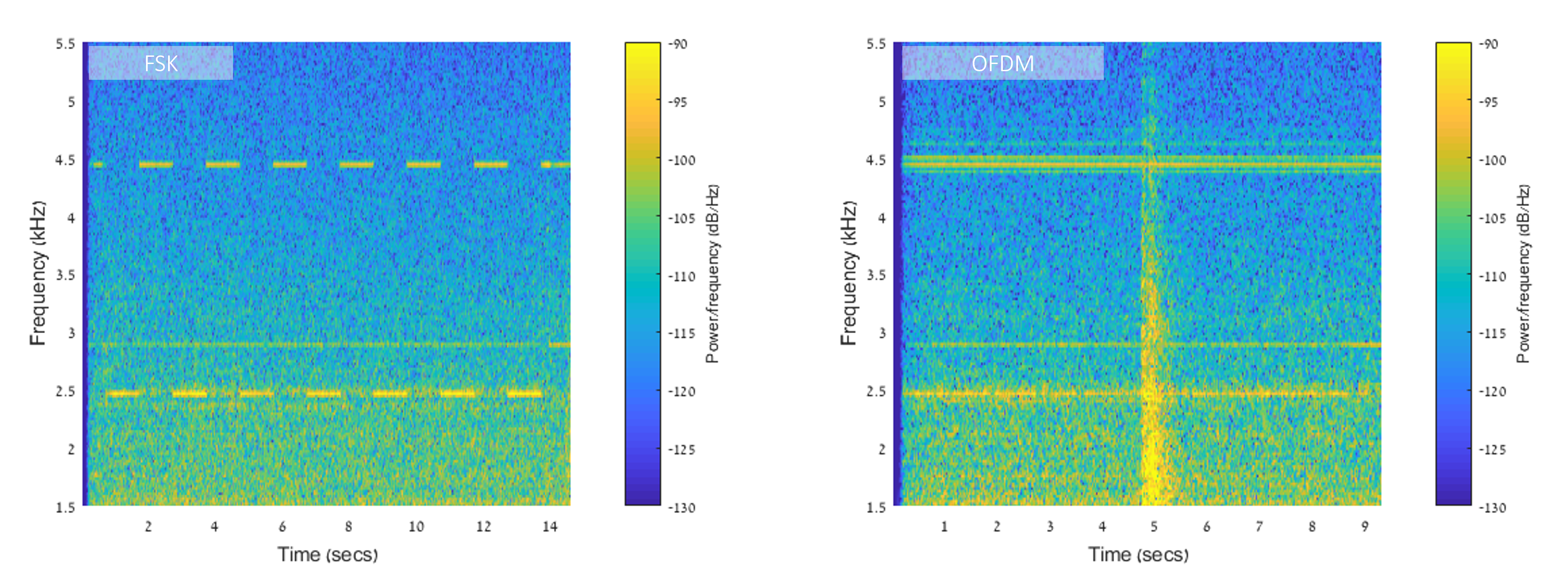}
	\caption{FSK (left) and OFDM (right).}
	\label{fig:multi}
\end{figure}

\begin{figure}
	\centering
	\includegraphics[width=\linewidth]{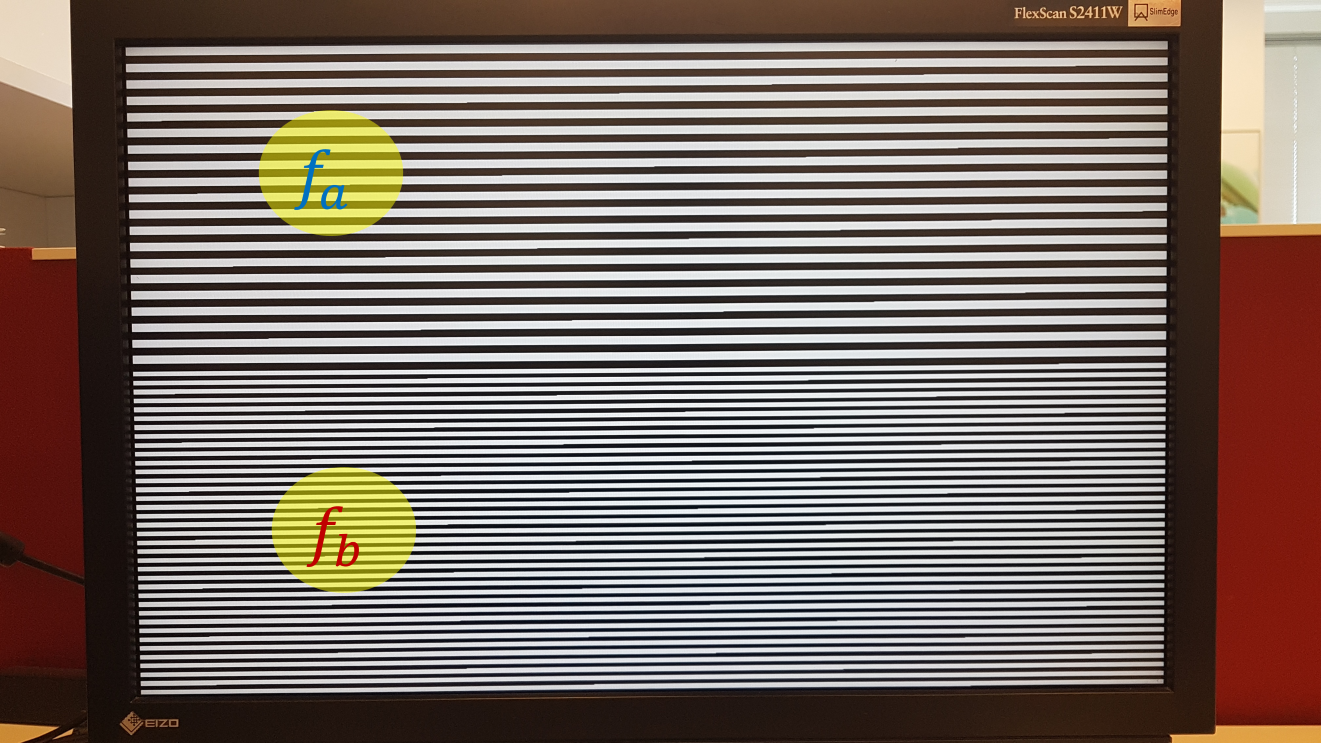}
	\caption{Splitted screen modulating two sub-carriers.}
	\label{fig:multi2}
\end{figure}

\subsection{Stealth and Brightness}
\label{sec:evasion}
In its standard form, the PIXELL attack is not hidden, i.e., it is visible to users looking at the LCD screen. To remain covert, attackers may use a strategy that transmits while the user is absent. For example, a so-called 'overnight attack' on the covert channels is maintained during the off-hours, reducing the risk of being revealed and exposed.

We developed an evasion technique for the PIXELL attack to remain stealthy during working hours. Our method is based on the assumption that low changes in brightness and pixel colors would be less noticeable to humans. In our techniques, the pixel colors are reduced to very low values before transmitting, which appears like a black screen to the unaware user who stands remotely. To achieve that effect, two techniques can be employed.

\begin{figure}
	\centering
	\includegraphics[width=\linewidth]{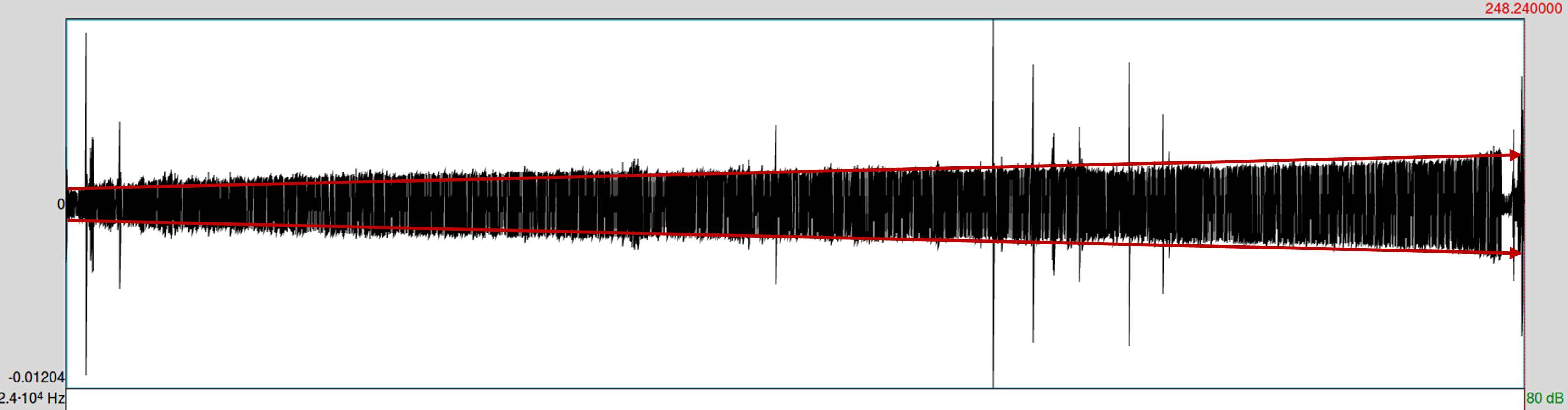}
	\caption{Waveform of the acoustic signal emanating from the LCD screen.}
	\label{fig:br1}
\end{figure}
\begin{figure}
	\centering
	\includegraphics[width=\linewidth]{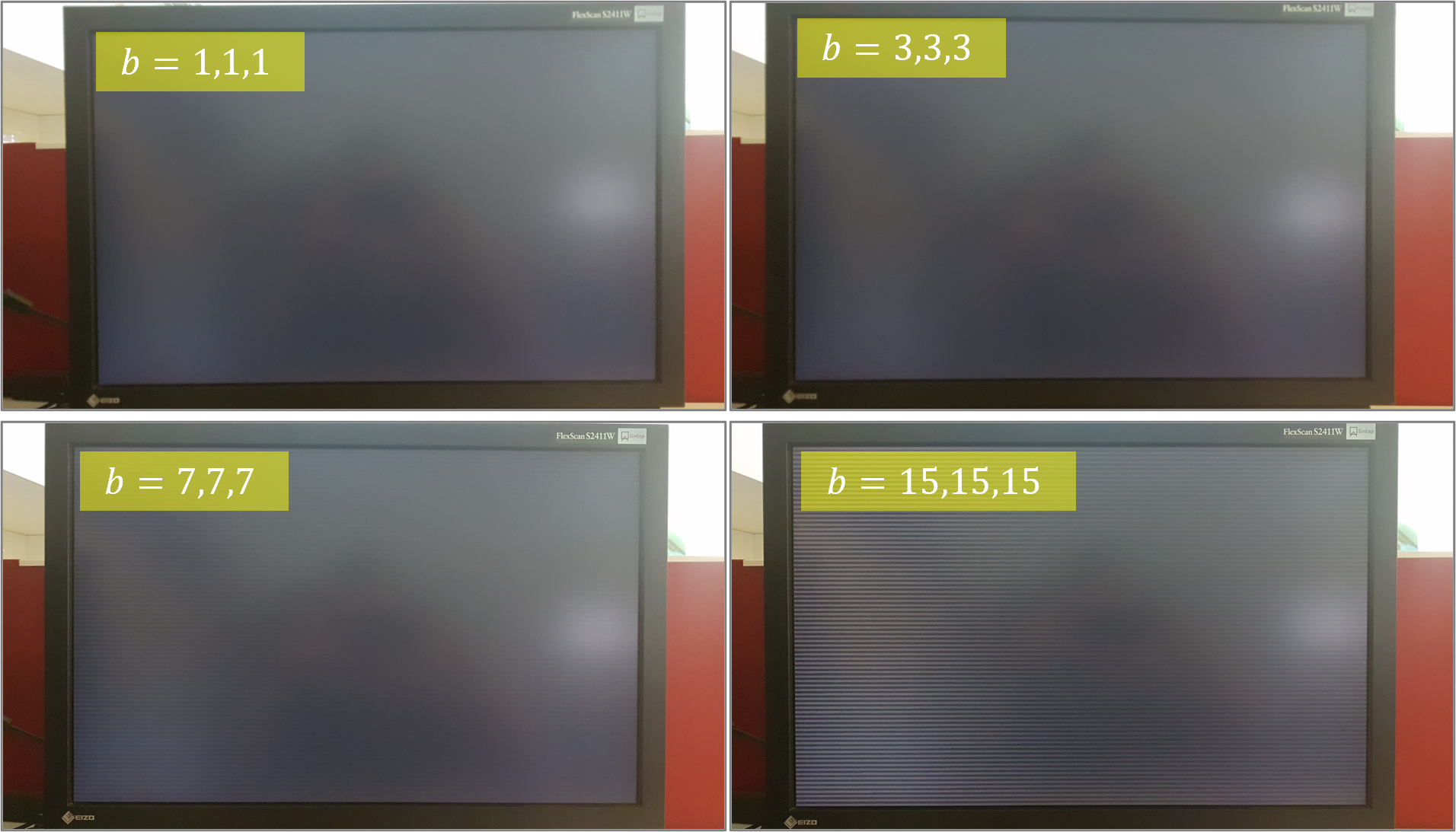}
	\caption{Transmitting screen with RGB levels of (1,1,1), (3,3,3), {7,7,7}, and {15,15,15}}
	\label{fig:br2}
\end{figure}

\begin{figure}
	\centering
	\includegraphics[width=0.8\linewidth]{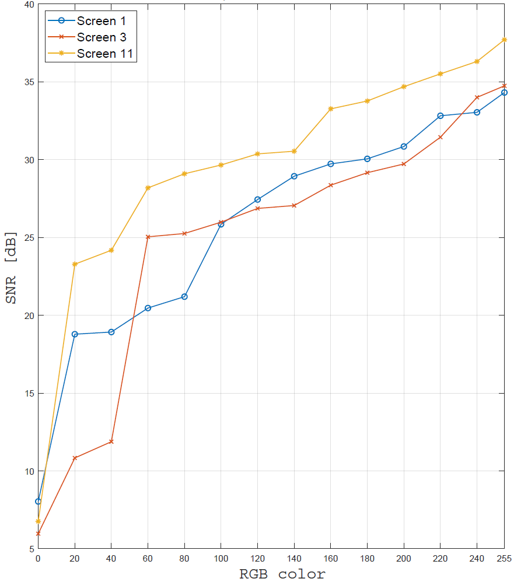}
	\caption{Measured SNR where the RGB colors are increased from (0,0,0) to (255,255,255) on three LCD screens.}
	\label{fig:snrb}
\end{figure}

\begin{itemize}
	\item \textbf{Backlight Control.}
	As mentioned in section \ref{sec:adv} backlight on LCD screens refers to a light source placed behind the liquid crystal layers to illuminate the display. LCD panels themselves do not produce light; they control light transmission through the liquid crystals to create images. In most cases, the backlight is placed directly behind the display. This configuration can offer better brightness uniformity and can support local dimming. This improves the contrast ratio by dimming the backlight in parts of the screen that display black while keeping it bright in lighter parts. Controlling the backlight of an LCD screen via software involves adjusting the brightness level of the screen's backlight to suit user preferences or ambient light conditions, usually in brightness values of 0-255. This control can be achieved through various methods and is supported by operating systems, device drivers, and specific applications. In Linux, the \texttt{xbacklight} command line tool can be used to set the backlight levels, e.g., \texttt{xbacklight -set 10} set the illumination levels to 10\%.
	
	\item \textbf{Pixels RGB Control.}
	The RGB values of the pixels shown on the screen are ($V$,$V$,$V$), with $V$ being 255 by default, which produces the most significant acoustic emission. By setting the RGB values of the pixels to lower values, it is possible to reduce the brightness of the generated bitmap. Figure \ref{fig:br2} shows a transmitting screen with RGB levels of (1,1,1), (3,3,3), (7,7,7), and (15,15,15).
\end{itemize}

It's important to note that brightness levels directly affect electrical current activity and, hence, the sound produced by the screen. Figure \ref{fig:br1} shows the waveform of the acoustic signal emanating from the LCD screen. Backlight levels are gradually increased from 10\% to 100\%.
However, lower RGB values yield lower emitted signals. This can be seen in Figure \ref{fig:snrb}, which shows the measured SNR where the RGB colors are graduated from (0,0,0) to (255,25,255) on three LCD screens. The acoustic signal's SNR has gradually increased from 4-6 dB to 33-37 dB, with the highest RGB values (255, 255, 255). It is important to note that this stealth technique is not bulletproof; if a user looks carefully at the screen, he can notice anomalous patterns. In addition, sound production is significantly reduced at lower brightness levels.

\begin{figure}
	\centering
	\includegraphics[width=\linewidth]{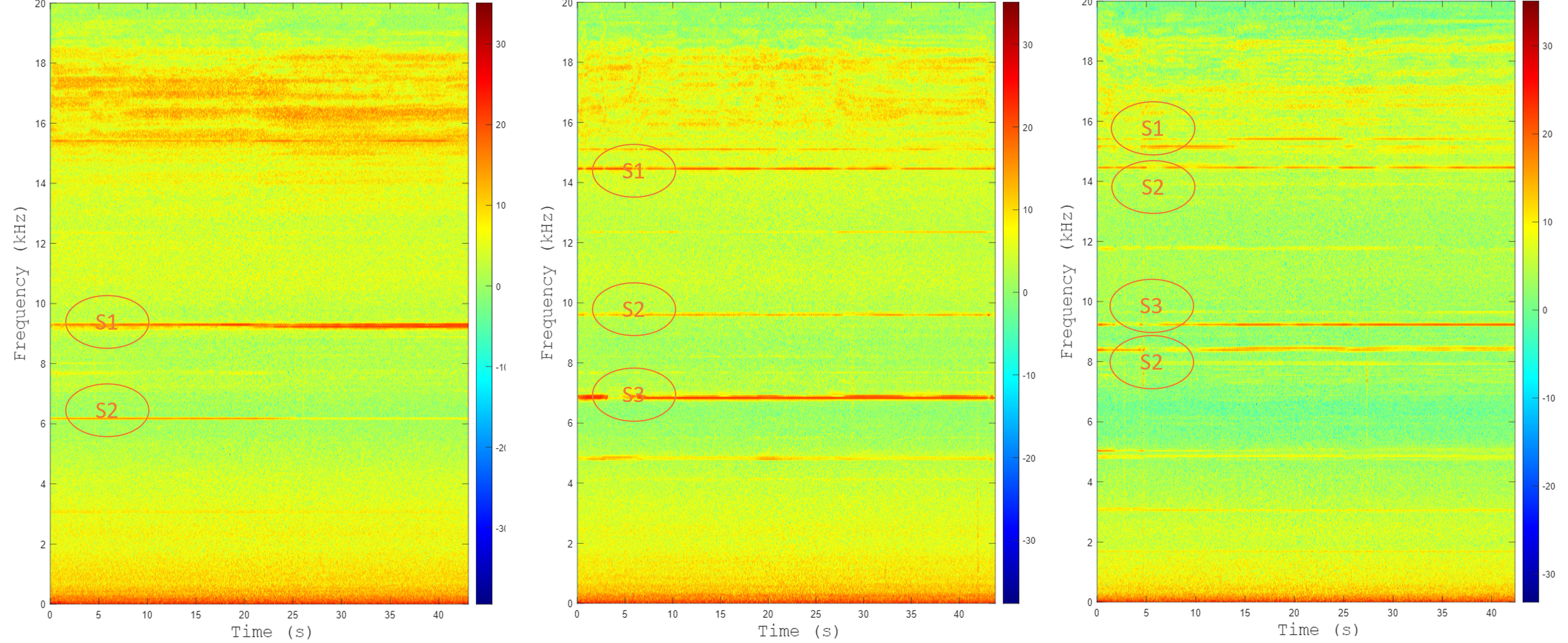}
	\caption{Spectrogram of signals received from one, two, three, and four screens simultaneously.}
	\label{fig:multis}
\end{figure}

\begin{figure}
	\centering
	\includegraphics[width=0.5\linewidth]{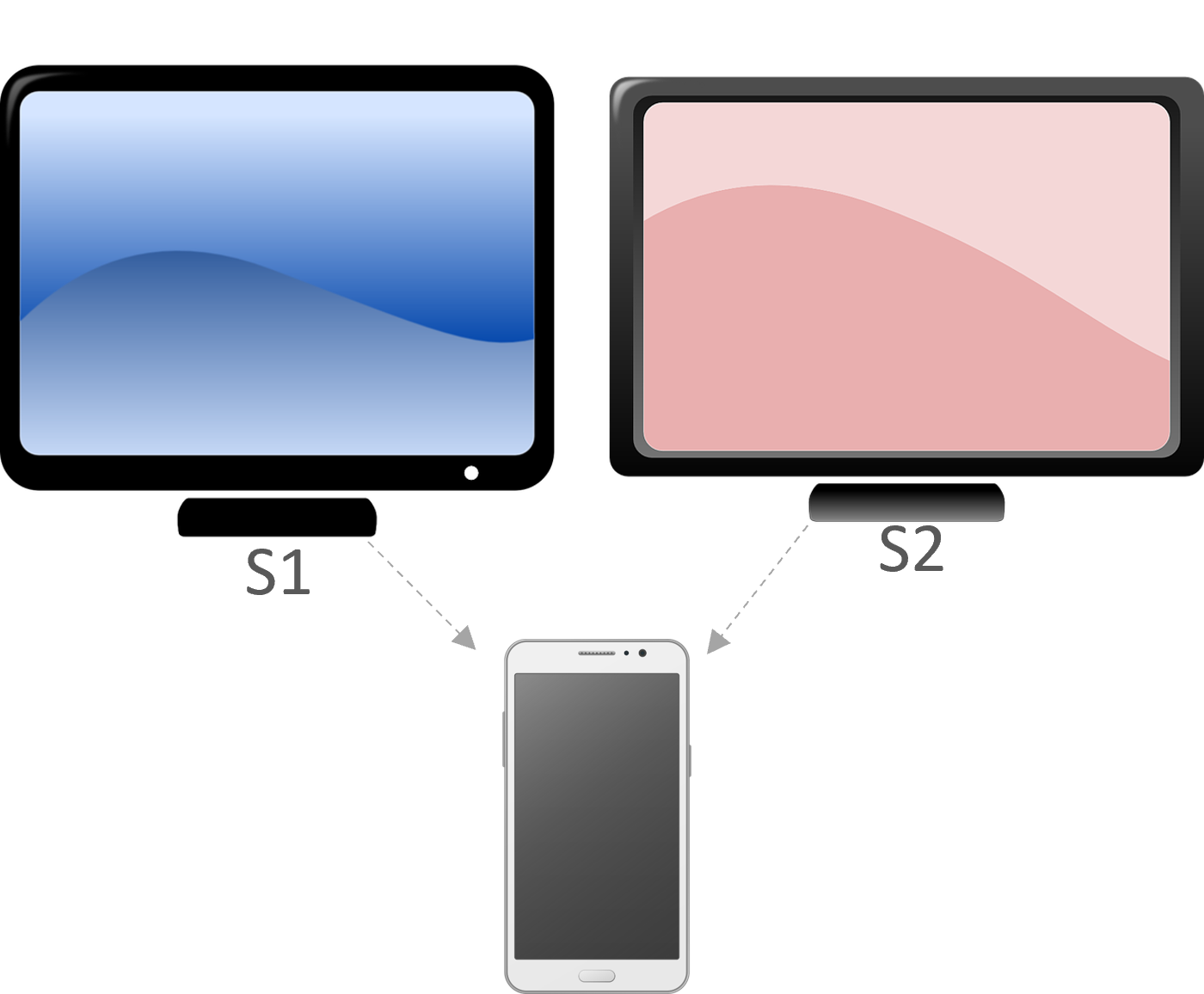}
	\caption{Maintaining a covert channel from two sources.}
	\label{fig:multisa}
\end{figure}

\subsection{Multiple Sources}
An attack scenario involving a single transmitter screen is the most common. Nevertheless, it is possible to maintain a covert channel from several sources as depicted in Figure \ref{fig:multis}. An example is when the user splits his working area across several screens. It is also possible to have two or more screens belonging to different users located on the same desk that belong to different users. As a result, malware can split data between several exfiltration points and leak it. To enable the receiver to aggregate the data correctly, additional meta information must be added to it. Based on $n$ transmitting screens, the bandwidth of the covert channels would be $BW * n$. It is important to choose the right carrier frequency for each transmitting computer to avoid collisions between the transmission band and its harmonics. Figure \ref{fig:multis} shows the spectrogram of signals received from one, two, three, and four screens simultaneously, in bands wrapping around 6 kHz, 9 kHz, 14.5 kHz, and 15.5 kHz. Notably, some signals are received stronger than others depending on where the screen is positioned and where the receiver is positioned.

\section{Countermeasures}
\label{sec:cnt}
Covert acoustic channels are countered by detecting, disrupting, or preventing their transmission.

\begin{itemize}
	\item \textbf{Jamming and Noise Generation.} With this defensive measure, ambient noise masking in the environment can interfere with the acoustic signals generated from the LCD screen, making transmission less reliable and harder to decode. Acoustic jammer devices work mainly on the principle of auditory masking, where complex noise obscures sound. They can be effective in various environments, from personal spaces to offices or meeting rooms, where the acoustic covert channel might occur. The PIXELL attack involves an equal-intensity random signal with a constant power spectrum density of 0-24 kHz. However, such devices' environmental and health implications can vary by jurisdiction, and they are considered less practical for wide deployment.
	
	\item \textbf{Spectrum Monitoring.} Monitoring the audio spectrum for unusual or uncommon signals can help detect the acoustic covert channel. This includes looking for data transmission patterns or frequencies outside the environmental norm. In environments with varying or high background noise levels, distinguishing between normal ambient sounds and covert transmissions can be challenging. Complex or fluctuating noise patterns can be identified as a covert channel, leading to false positives \cite{carrara2014acoustic}. Note that the PIXHELL attack might use frequencies that overlap with legitimate audio sources. For example, the ultrasonic band might share frequencies with items that emit ultrasonic noise during normal operation, such as power supplies and other electronic devices in the area.
	
	\item \textbf{Access Control and Physical Security.} Limiting physical access to sensitive areas and ensuring that devices that could be used to initiate or receive acoustic signals are kept out of these areas can reduce the risk of acoustic covert channels being established. Banning smartphones, laptops, and microphones from sensitive areas may reduce data exfiltration risk.
	
	\item \textbf{Anomaly Detection Systems.} Implementing anomaly detection systems that identify suspicious behavior on the bitmap shown on the screen. In the case of a PIXELL attack, an analysis of the screen buffer can be used to detect the transmission pattern, which usually consists of white pixels on iterative patterns that encode a specific frequency. Such a solution can be implemented at the user or kernel level. For example, utilizing GDI functions to capture the screen. In Windows OS, Win32 API functions such as \texttt{BitBlt()} \cite{BitBltfu33:online} can be used to copy the screen buffer or a window into a bitmap. Such security solutions can be bypassed by malware running on a compromised computer at the user and kernel levels.
	
	\item \textbf{External camera monitoring.} Detecting the covert channel using an external camera and image analysis applies to monitor for unusual modulated screen patterns. It involves developing algorithms for specific patterns indicative of data transmission, such as white pixel bitmaps with certain intervals, intensity changes, or color shifts. However, this solution is not always feasible since introducing monitoring cameras to specific places might pose a security challenge.
\end{itemize}

\section{Conclusion}
\label{sec:conclude}
This paper presents PIXHELL, an acoustic covert channel on air-gap, audio-gapped computers. We show that although these systems lack standard audio capabilities (e.g., loudspeakers), attackers can abuse the screen to generate sound. Our technique uses special pixel patterns to control the sound emission from the screen's internal electric components due to coil whining and piezoelectric effects. We presented the attack model and discussed scenarios such as malicious insiders and compromised mobile devices. We compared this work with others in the domain and presented its contribution. We provided the design and implementation of a transmitter and receiver on mobile and laptop and discussed various modulation methods, such as M-FSK and OFDM. We evaluate the covert channel in several aspects, including distance and speed. We also proposed a stealth and evasion technique that uses low-brightness bitmaps to conceal the covert channel. Finally, we provide defensive countermeasures to the PIXHELL attack. Our results showed that various data types can be exfiltrated to nearby mobile phones, microphones, or laptops at a distance of 2m or more.

\balance      
\bibliographystyle{plain}

\begin{thebibliography}{10}
	
	\bibitem{AirGapCo10:online}
	Air gap computer network security - notary colorado springs.
	\newblock
	\url{https://abclegaldocs.com/blog-Colorado-Notary/air-gap-computer-network-security/}.
	\newblock (Accessed on 02/28/2024).
	
	\bibitem{America10:online}
	Americas electric grid has a vulnerable back door—and russia walked through
	it - wsj.
	\newblock
	\url{https://www.wsj.com/articles/americas-electric-grid-has-a-vulnerable-back-doorand-russia-walked-through-it-11547137112}.
	\newblock (Accessed on 02/28/2024).
	
	\bibitem{Beatingt16:online}
	Beating the air-gap: How attackers can gain access to supposedly isolated
	systems | energy central.
	\newblock
	\url{https://www.energycentral.com/c/iu/beating-air-gap-how-attackers-can-gain-access-supposedly-isolated-systems}.
	\newblock (Accessed on 01/01/2023).
	
	\bibitem{BitBltfu33:online}
	Bitblt function (wingdi.h) - win32 apps | microsoft learn.
	\newblock
	\url{https://learn.microsoft.com/en-us/windows/win32/api/wingdi/nf-wingdi-bitblt}.
	\newblock (Accessed on 02/28/2024).
	
	\bibitem{ChinasAP62:online}
	China's apt31 suspected in attacks on air-gapped systems in eastern europe.
	\newblock
	\url{https://thehackernews.com/2023/08/chinas-apt31-suspected-in-attacks-on.html}.
	\newblock (Accessed on 12/31/2023).
	
	\bibitem{Electrom69:online}
	Electromagnetically induced acoustic noise - wikipedia.
	\newblock
	\url{https://en.wikipedia.org/wiki/Electromagnetically_induced_acoustic_noise}.
	\newblock (Accessed on 02/28/2024).
	
	\bibitem{USBDevic57:online}
	Usb devices the common denominator in all attacks on air-gapped systems.
	\newblock
	\url{https://www.darkreading.com/cyberattacks-data-breaches/usb-devices-common-denominator-in-all-attacks-on-air-gapped-systemsd}.
	\newblock (Accessed on 02/28/2024).
	
	\bibitem{WaybackM84:online}
	Wayback machine.
	\newblock
	\url{https://web.archive.org/web/20131203022542/http://www.insurancenewsnet.com/images/post/112612_Weber_complaint.pdf}.
	\newblock (Accessed on 12/30/2023).
	
	\bibitem{carrara2016air}
	Brent Carrara.
	\newblock {\em Air-Gap Covert Channels}.
	\newblock PhD thesis, Universit{\'e} d'Ottawa/University of Ottawa, 2016.
	
	\bibitem{carrara2014acoustic}
	Brent Carrara and Carlisle Adams.
	\newblock On acoustic covert channels between air-gapped systems.
	\newblock In {\em International Symposium on Foundations and Practice of
		Security}, pages 3--16. Springer, 2014.
	
	\bibitem{carrara2015acoustic}
	Brent Carrara and Carlisle Adams.
	\newblock On acoustic covert channels between air-gapped systems.
	\newblock In {\em Foundations and Practice of Security: 7th International
		Symposium, FPS 2014, Montreal, QC, Canada, November 3-5, 2014. Revised
		Selected Papers 7}, pages 3--16. Springer, 2015.
	
	\bibitem{clark2009hardware}
	John Clark, Sylvain Leblanc, and Scott Knight.
	\newblock Hardware trojan horse device based on unintended usb channels.
	\newblock In {\em Network and System Security, 2009. NSS'09. Third
		International Conference on}, pages 1--8. IEEE, 2009.
	
	\bibitem{cronin2019covert}
	Patrick Cronin, Charles Gouert, Dimitris Mouris, Nektarios~Georgios Tsoutsos,
	and Chengmo Yang.
	\newblock Covert data exfiltration using light and power channels.
	\newblock In {\em 2019 IEEE 37th International Conference on Computer Design
		(ICCD)}, pages 301--304. IEEE, 2019.
	
	\bibitem{de2022inkfiltration}
	Julian de~Gortari~Briseno, Akash~Deep Singh, and Mani Srivastava.
	\newblock Inkfiltration: Using inkjet printers for acoustic data exfiltration
	from air-gapped networks.
	\newblock {\em ACM Transactions on Privacy and Security}, 25(2):1--26, 2022.
	
	\bibitem{deshotels2014inaudible}
	Luke Deshotels.
	\newblock Inaudible sound as a covert channel in mobile devices.
	\newblock {\em WOOT}, 14:16--1, 2014.
	
	\bibitem{dorais2021jumping}
	Alexis Dorais-Joncas and Facundo Mun{\~o}z.
	\newblock Jumping the air gap.
	\newblock 2021.
	
	\bibitem{faou2020agent}
	Matthieu Faou.
	\newblock From agent. btz to comrat v4, 2020.
	
	\bibitem{guri2020cd}
	Mordechai Guri.
	\newblock Cd-leak: Leaking secrets from audioless air-gapped computers using
	covert acoustic signals from cd/dvd drives.
	\newblock In {\em 2020 IEEE 44th Annual Computers, Software, and Applications
		Conference (COMPSAC)}, pages 808--816. IEEE, 2020.
	
	\bibitem{guri2021exfiltrating}
	Mordechai Guri.
	\newblock Exfiltrating data from air-gapped computers via vibrations.
	\newblock {\em Future Generation Computer Systems}, 2021.
	
	\bibitem{guri2021gairoscope}
	Mordechai Guri.
	\newblock Gairoscope: Leaking data from air-gapped computers to nearby
	smartphones using speakers-to-gyro communication.
	\newblock In {\em 2021 18th International Conference on Privacy, Security and
		Trust (PST)}, pages 1--10. IEEE, 2021.
	
	\bibitem{GURI2021115}
	Mordechai Guri.
	\newblock Magneto: Covert channel between air-gapped systems and nearby
	smartphones via cpu-generated magnetic fields.
	\newblock {\em Future Generation Computer Systems}, 115:115 -- 125, 2021.
	
	\bibitem{guri2021power}
	Mordechai Guri.
	\newblock Power-supplay: Leaking sensitive data from air-gapped, audio-gapped
	systems by turning the power supplies into speakers.
	\newblock {\em IEEE Transactions on Dependable and Secure Computing}, 2021.
	
	\bibitem{guri2021usbculprit}
	Mordechai Guri.
	\newblock Usbculprit: Usb-borne air-gap malware.
	\newblock In {\em European Interdisciplinary Cybersecurity Conference}, pages
	7--13, 2021.
	
	\bibitem{guri2022gpu}
	Mordechai Guri.
	\newblock Gpu-fan: Leaking sensitive data from air-gapped machines via covert
	noise from gpu fans.
	\newblock In {\em Nordic Conference on Secure IT Systems}, pages 194--211.
	Springer, 2022.
	
	\bibitem{guri2023rambo}
	Mordechai Guri.
	\newblock Rambo: Leaking secrets from air-gap computers by spelling covert
	radio signals from computer ram.
	\newblock In {\em Nordic Conference on Secure IT Systems}, pages 144--161.
	Springer, 2023.
	
	\bibitem{guri2019brightness}
	Mordechai Guri, Dima Bykhovsky, and Yuval Elovici.
	\newblock Brightness: Leaking sensitive data from air-gapped workstations via
	screen brightness.
	\newblock In {\em 2019 12th CMI Conference on Cybersecurity and Privacy (CMI)},
	pages 1--6. IEEE, 2019.
	
	\bibitem{Guri:2018:BAM:3200906.3177230}
	Mordechai Guri and Yuval Elovici.
	\newblock Bridgeware: The air-gap malware.
	\newblock {\em Commun. ACM}, 61(4):74--82, March 2018.
	
	\bibitem{guri2015gsmem}
	Mordechai Guri, Assaf Kachlon, Ofer Hasson, Gabi Kedma, Yisroel Mirsky, and
	Yuval Elovici.
	\newblock Gsmem: Data exfiltration from air-gapped computers over gsm
	frequencies.
	\newblock In {\em USENIX Security Symposium}, pages 849--864, 2015.
	
	\bibitem{guri2014airhopper}
	Mordechai Guri, Gabi Kedma, Assaf Kachlon, and Yuval Elovici.
	\newblock Airhopper: Bridging the air-gap between isolated networks and mobile
	phones using radio frequencies.
	\newblock In {\em Malicious and Unwanted Software: The Americas (MALWARE), 2014
		9th International Conference on}, pages 58--67. IEEE, 2014.
	
	\bibitem{guri2015bitwhisper}
	Mordechai Guri, Matan Monitz, Yisroel Mirski, and Yuval Elovici.
	\newblock Bitwhisper: Covert signaling channel between air-gapped computers
	using thermal manipulations.
	\newblock In {\em Computer Security Foundations Symposium (CSF), 2015 IEEE
		28th}, pages 276--289. IEEE, 2015.
	
	\bibitem{guri2017acoustic}
	Mordechai Guri, Yosef Solewicz, Andrey Daidakulov, and Yuval Elovici.
	\newblock Acoustic data exfiltration from speakerless air-gapped computers via
	covert hard-drive noise (diskfiltration).
	\newblock In {\em European Symposium on Research in Computer Security}, pages
	98--115. Springer, 2017.
	
	\bibitem{Guri2017e}
	Mordechai Guri, Yosef Solewicz, Andrey Daidakulov, and Yuval Elovici.
	\newblock Speake(a)r: Turn speakers to microphones for fun and profit.
	\newblock In {\em 11th USENIX Workshop on Offensive Technologies (WOOT 17)}.
	USENIX Association, 2017.
	
	\bibitem{guri2018mosquito}
	Mordechai Guri, Yosef Solewicz, and Yuval Elovici.
	\newblock Mosquito: Covert ultrasonic transmissions between two air-gapped
	computers using speaker-to-speaker communication.
	\newblock In {\em 2018 IEEE Conference on Dependable and Secure Computing
		(DSC)}, pages 1--8. IEEE, 2018.
	
	\bibitem{guri2020fansmitter}
	Mordechai Guri, Yosef Solewicz, and Yuval Elovici.
	\newblock Fansmitter: Acoustic data exfiltration from air-gapped computers via
	fans noise.
	\newblock {\em Computers \& Security}, page 101721, 2020.
	
	\bibitem{guri2019ctrl}
	Mordechai Guri, Boris Zadov, Dima Bykhovsky, and Yuval Elovici.
	\newblock Ctrl-alt-led: Leaking data from air-gapped computers via keyboard
	leds.
	\newblock In {\em 2019 IEEE 43rd Annual Computer Software and Applications
		Conference (COMPSAC)}, volume~1, pages 801--810. IEEE, 2019.
	
	\bibitem{guri2019powerhammer}
	Mordechai Guri, Boris Zadov, Dima Bykhovsky, and Yuval Elovici.
	\newblock Powerhammer: Exfiltrating data from air-gapped computers through
	power lines.
	\newblock {\em IEEE Transactions on Information Forensics and Security}, 2019.
	
	\bibitem{guri2019odini}
	Mordechai Guri, Boris Zadov, and Yuval Elovici.
	\newblock Odini: Escaping sensitive data from faraday-caged, air-gapped
	computers via magnetic fields.
	\newblock {\em IEEE Transactions on Information Forensics and Security},
	15:1190--1203, 2019.
	
	\bibitem{hanspach2014covert}
	Michael Hanspach and Michael Goetz.
	\newblock On covert acoustical mesh networks in air.
	\newblock {\em arXiv preprint arXiv:1406.1213}, 2014.
	
	\bibitem{karnouskos2011stuxnet}
	Stamatis Karnouskos.
	\newblock Stuxnet worm impact on industrial cyber-physical system security.
	\newblock In {\em IECON 2011-37th Annual Conference of the IEEE Industrial
		Electronics Society}, pages 4490--4494. IEEE, 2011.
	
	\bibitem{martinez2021software}
	Jeferson Martinez and Javier~M Duran.
	\newblock Software supply chain attacks, a threat to global cybersecurity:
	Solarwinds’ case study.
	\newblock {\em International Journal of Safety and Security Engineering},
	11(5):537--545, 2021.
	
	\bibitem{masti2015thermal}
	Ramya~Jayaram Masti, Devendra Rai, Aanjhan Ranganathan, Christian M{\"u}ller,
	Lothar Thiele, and Srdjan Capkun.
	\newblock Thermal covert channels on multi-core platforms.
	\newblock In {\em 24th $\{$USENIX$\}$ Security Symposium ($\{$USENIX$\}$
		Security 15)}, pages 865--880, 2015.
	
	\bibitem{matyunin2016covert}
	Nikolay Matyunin, Jakub Szefer, Sebastian Biedermann, and Stefan Katzenbeisser.
	\newblock Covert channels using mobile device's magnetic field sensors.
	\newblock In {\em Design Automation Conference (ASP-DAC), 2016 21st Asia and
		South Pacific}, pages 525--532. IEEE, 2016.
	
	\bibitem{matyunin2019vibrational}
	Nikolay Matyunin, Yujue Wang, and Stefan Katzenbeisser.
	\newblock Vibrational covert channels using low-frequency acoustic signals.
	\newblock In {\em Proceedings of the ACM Workshop on Information Hiding and
		Multimedia Security}, pages 31--36, 2019.
	
	\bibitem{pandya2022shoutimei}
	Keval Pandya, Bhavesh Borisaniya, and Bharat Buddhadev.
	\newblock Shoutimei: Ultrasound covert channel-based attack in android.
	\newblock In {\em Security, Privacy and Data Analytics: Select Proceedings of
		ISPDA 2021}, pages 293--301. Springer, 2022.
	
	\bibitem{shen2021lora}
	Cheng Shen, Tian Liu, Jun Huang, and Rui Tan.
	\newblock When lora meets emr: Electromagnetic covert channels can be super
	resilient.
	\newblock In {\em 2021 IEEE Symposium on Security and Privacy (SP)}, pages
	1304--1317. IEEE, 2021.
	
	\bibitem{sherry2023near}
	R~Sherry, E~Bayne, and D~McLuskie.
	\newblock Near-ultrasonic covert channels using software-defined radio
	techniques.
	\newblock In {\em Proceedings of the International Conference on Cybersecurity,
		Situational Awareness and Social Media: Cyber Science 2022; 20--21 June;
		Wales}, pages 169--189. Springer, 2023.
	
	\bibitem{wong2018crossing}
	Wesley Wong et~al.
	\newblock {\em Crossing the air gap—an ultrasonic covert channel}.
	\newblock PhD thesis, 2018.
	
	\bibitem{zhan2020bitjabber}
	Zihao Zhan, Zhenkai Zhang, and Xenofon Koutsoukos.
	\newblock Bitjabber: The world’s fastest electromagnetic covert channel.
	\newblock In {\em 2020 IEEE International Symposium on Hardware Oriented
		Security and Trust (HOST)}, pages 35--45. IEEE, 2020.
	
	\bibitem{zhang2022ultrannel}
	Jianyi Zhang, Ruilong Wu, Xiuying Li, Shuo Wang, Qichao Jin, Zhenkui Li, and
	Rui Shi.
	\newblock Ultrannel: Ultrasound based covert communication channel.
	\newblock In {\em 2022 IEEE Smartworld, Ubiquitous Intelligence \& Computing,
		Scalable Computing \& Communications, Digital Twin, Privacy Computing,
		Metaverse, Autonomous \& Trusted Vehicles
		(SmartWorld/UIC/ScalCom/DigitalTwin/PriComp/Meta)}, pages 1760--1767. IEEE,
	2022.
	
	\bibitem{zhang2020magview}
	Juchuan Zhang, Xiaoyu Ji, Wenyuan Xu, Yi-Chao Chen, Yuting Tang, and Gang Qu.
	\newblock Magview: A distributed magnetic covert channel via video encoding and
	decoding.
	\newblock In {\em IEEE INFOCOM 2020-IEEE Conference on Computer
		Communications}, pages 357--366. IEEE, 2020.
	
\end{thebibliography}

\end{document}